# New R Coronae Borealis and DY Persei Star Candidates and Other Related Objects Found in Photometric Surveys

**Sebastián Otero**
*AAVSO Headquarters, 49 Bay State Road, Cambridge, MA 02138; address email correspondence to sebastian@aavso.org*

**Stefan Hümmerich**
*Stiftstr. 4, Braubach, D-56338, Germany*

**Klaus Bernhard**
*Kafkaweg 5, Linz, 4030, Austria*

**Igor Soszyński**
*Warsaw University Observatory, Al. Ujazdowskie 4, 00-478 Warszawa, Poland*

**Abstract** We have carried out a search for new R Coronae Borealis (RCB) variables using the publicly accessible data from various photometric sky surveys and—whenever available—AAVSO visual data. Candidates were selected from Tisserand's "Catalogue enriched with R CrB stars" and by a visual inspection of light curves from the ASAS-3, MACHO, NSVS and OGLE surveys. We have identified two new RCB stars, four RCB candidates, and one DY Persei (DYPer) star candidate. Our identification was based mainly on photometric variability, color-color diagrams, and further information drawn from various catalogue sources; spectroscopic classifications were also reported in our analysis whenever available. Additionally, we present a sample of interesting stars which—although showing similar photometric variability—can be ruled out as RCB and DYPer stars or have been rejected as such on spectroscopic grounds in recent studies. Although not useful in the investigation of the aforementioned groups of variables, these objects defy an easy classification and might be interesting targets for follow-up studies which we encourage for all stars presented in this paper.

## 1. Introduction

R Coronae Borealis (hereafter RCB) stars are a rare class of variables characterized by peculiar chemical composition (notably hydrogen deficiency and carbon overabundance) and unusual photometric variability (irregular and unpredictable fading events). They are a poorly understood class of variables and controversy about their origin is still going on, with recent evidence favoring the Double Degenerate (DD) over the Final Helium Shell Flash (FF) scenario (see, for example, Clayton (2012)).

In order to better understand these enigmatic objects, it is imperative to increase the sample of known RCB stars. With the advent of photometric surveys, progress has been made in this respect. Furthermore, it has been shown that near- and mid-infrared color-color diagrams and cuts are a viable and efficient method of identifying new RCB candidates (see, for example, Feast (1997); Alcock *et al.* (2001); Tisserand *et al.* (2004); and Tisserand (2012)). At the time of this writing, the number of confirmed RCB stars has increased to 76 galactic and 22 extragalactic objects (Tisserand *et al.* 2013b).

Among the RCB variables, DY Persei (hereafter DYPer) stars have been of special interest for being situated at the lower end of the temperature scale. Recently, however, evidence has been mounting that DYPer stars might have more in common with ordinary carbon stars than with other RCB variables (see, for example, Tisserand *et al.* (2009) and Soszyński *et al.* (2009)). In addition to their lower temperatures, they generally show slower declines with smaller amplitudes and roughly symmetric recoveries. Furthermore, they are fainter on average and their pulsational periods tend to be longer than those of typical RCB stars. They are further set aside by a relatively high amount of $^{13}C$ in their spectra—an isotope of carbon, whose shortage or absence is one of the defining characteristics of classical RCB stars (see, for example, Lloyd Evans (2010)). However, the characterization of the DYPer stars still suffers from an insufficient sample size. Increasing the number of known DYPer stars is therefore an important task.

This paper presents two new RCB stars, four RCB candidates, and one DYPer candidate that have been found using data from various photometric surveys. To achieve this, two different approaches were taken. Firstly, candidates from the VizieR online version of the "Catalogue enriched with R CrB stars" (Tisserand 2012) were investigated using data from various sky surveys and catalogues. Secondly, preselected light curves from the ASAS-3 (Pojmański 2002), MACHO (Alcock *et al.* 1997), NSVS (Woźniak *et al.* 2004b), and OGLE (Udalski *et al.* 1997) surveys were inspected visually for stars showing conspicuous fading events. It is important to note that, for the present paper, the identification of a star as a possible RCB or DYPer candidate has been based primarily on the object's photometric behavior. Further information drawn from color-color diagrams and various catalogue sources has also been included in the analysis.

Additionally, we present a choice sample of interesting stars which—although showing similar photometric variability—can be ruled out as RCB and DYPer stars or have been rejected as such on spectroscopic grounds in recent studies. Although not useful in the investigation of the aforementioned groups of variables, these objects defy an easy classification and might be interesting targets for follow-up studies.

## 2. Target stars

### 2.1. Overview

An overview providing essential data of all variables studied in this paper is given in Table 1. Objects have been sorted by proposed type (RCB / RCB: / DYPer: / SR:) and then by right ascension. Each object is discussed in detail below. For a general definition of variability types, refer to the Variable Star Type Designations in VSX (Otero *et al.* 2013), which are based on the *General Catalogue of Variable Stars* (GCVS; Samus *et al.* 2007–2013) variability types documentation with expansions and revisions from the literature.

### 2.2. Near- and mid-infrared color-color diagrams

Color-color diagrams have been successfully employed in the investigation of RCB variables (see, for example, Feast (1997), Alcock *et al.* (2001), Tisserand *et al.* (2004), and Tisserand (2012)). We have used near- and mid-infrared color-color diagrams, based on data from the 2MASS and WISE (Cutri *et al.* 2012) catalogues, to investigate all stars presented in this paper.

#### 2.2.1. (J–H) vs. (H–K) diagram

As dust forms and disperses, the colors of RCB stars change. Feast (1997) employed (J–H) vs. (H–K) diagrams and found them to be valuable tools in the investigation of RCB color evolution. Furthermore, they can be used to effectively identify RCB stars and discriminate them from the related DYPer variables (see, for example, Alcock *et al.* (2001) and Tisserand *et al.* (2004)).

A (J–H) vs. (H–K) diagram for all stars presented in this paper is shown in Figure 1. The solid line indicates the colors of SMC carbon stars, as computed by Westerlund *et al.* (1991) and employed in this particular context by Tisserand *et al.* (2004). The dashed line indicates the loci of a combination of two blackbodies, representing the photosphere of the star (~5500 K) and the dust shell (~900 K), as devised by Feast (1997). The flux ranges from "all star" (lower end) to "all shell" (upper end).

As becomes obvious from Figure 1, the present sample is separated into two distinct groups. The two RCB stars and the RCB candidates (denoted by colored squares) roughly follow the dashed line that outlines their possible range in color that is due to the amount of circumstellar dust. The DYPer candidate, on the other hand, is situated near the expected loci of classical carbon stars, which also holds true for the set of miscellaneous variables which have been shown to undergo significant fading events but have been ruled out as RCB or DYPer stars on various grounds (section 3). The displacement of OGLE-II BUL-SC18 64562 towards redder colors might be explained by the heavy extinction in this specific region towards the Galactic Bulge.

Table 1. Essential data on all variables studied in this paper, sorted by proposed type and right ascension.

| *Identifiers* | *R.A. (J2000)* | *Dec. (J2000)* | *Type* | *Range* | *Period* | *Remarks* |
|---|---|---|---|---|---|---|
| ASAS J050232–7218.9<br>OGLE-IV LMC507.21.5<br>GSC 09169-00810<br>2MASS J05023226–7218534 | 05 02 32.270 | –72 18 53.58 | RCB | 13.5–<18.1 V | 43.1 d | In the LMC, Spectral type C2,2Hd.<br>J–K = 0.46; B–V = 1.34. |
| AO Her<br>IRAS 17343+5026<br>2MASS J17353628+5024398 | 17 35 36.284 | +50 24 39.88 | RCB | 10.7–<19.6: V | | Spectroscopically confirmed.<br>J–K = 3.59; B–V = 1.86. |
| NSVS J0051273+645649<br>GSC 04025-00779<br>IRAS 00483+6440 | 00 51 28.114 | +64 56 51.72 | RCB: | 13.2–15.0: V | 29.8 d | J–K = 1.76; B–V = 1.48. |
| IRAS 04519+3553<br>2MASS J04552045+3558079 | 04 55 20.456 | +35 58 07.98 | RCB: | >12.5–<16.0 CV | 438 d | J–K = 4.18. Large amplitude pulsations? |
| IZ Sgr<br>GSC 06279-00870<br>IRAS 18335–2101 | 18 36 31.256 | –20 59 15.49 | RCB: | 12.2–<17.0 V<br>13.3–19.1 B<br>>12.9–<17.6 R | | Reported spectral type of M6 erroneous?<br>J–K = 3.00. |
| NSVS 1461135<br>GSC 04282-00656<br>IRAS 23004+6300 | 23 02 28.518 | +63 16 31.04 | RCB: | 12.8–14.2 R1 | | J–K = 1.47; B–V = 0.63. |

*Table continued on next page*

Table 1. Essential data on all variables studied in this paper, cont.

| *Identifiers* | *R.A. (J2000)* | *Dec. (J2000)* | *Type* | *Range* | *Period* | *Remarks* |
|---|---|---|---|---|---|---|
| MACHO 128.21543.435 | 18 06 31.549 | –28 34 30.14 | DYPer: | 17.9–20.5: V | 75.6 d | J–K = 1.71. |
| EROS2–cg6143l34028 | | | | 15.0–17.15 $R_c$ | 66.4 d | |
| 2MASS J18063154–2834301 | | | | 12.7–<14.4 $I_c$ | | |
| ASAS J095221–4329.8 | 09 52 21.381 | –43 29 40.52 | SR: | 10.4–13.0 V | ~65 d | Spectral type M6; |
| CD–42 5700 | | | | | | “strong TiO, VO; H?” |
| IRAS 09503–4315 | | | | | | J–K = 1.29; B–V = 1.78. |
| ASAS J123034–7703.9 | 12 30 34.218 | –77 03 52.75 | SR: | 10.0–12.7 V | ~273 d | Si emission at 9.7 μm. |
| GSC 09416-00380 | | | | | | J–K = 1.67, B–V = 2.26. |
| IRAS 12274–7647 | | | | | | |
| OGLE-II BUL-SC18 64562 | 18 06 46.557 | –27 22 06.31 | SR: | 13.4:–15.8: $I_c$ | ~60 d | J–K = 2.47, V–I = 4.92. |
| OGLE-BLG-LPV-210930 | | | | | | |
| OGLE-IV BLG518.06.142574 | | | | | | |
| 2MASS J18064655–2722063 | | | | | | |
| NSV 12817 | 20 09 03.882 | –49 41 25.57 | SR: | 10.5 – 12.3 V | 50.93 d | J–K = 1.20, B–V = 1.58. |
| CD–50 12825 | | | | | | |
| IRAS 20054–4950 | | | | | | |

*Note: Positional data were taken from 2MASS (Skrutskie et al. 2006; IRAS 04519+3553, IZ Sgr, MACHO 128.21543.435, OGLE–II BUL–SC18 64562) and UCAC4 (Zacharias et al. 2012; all other objects).*

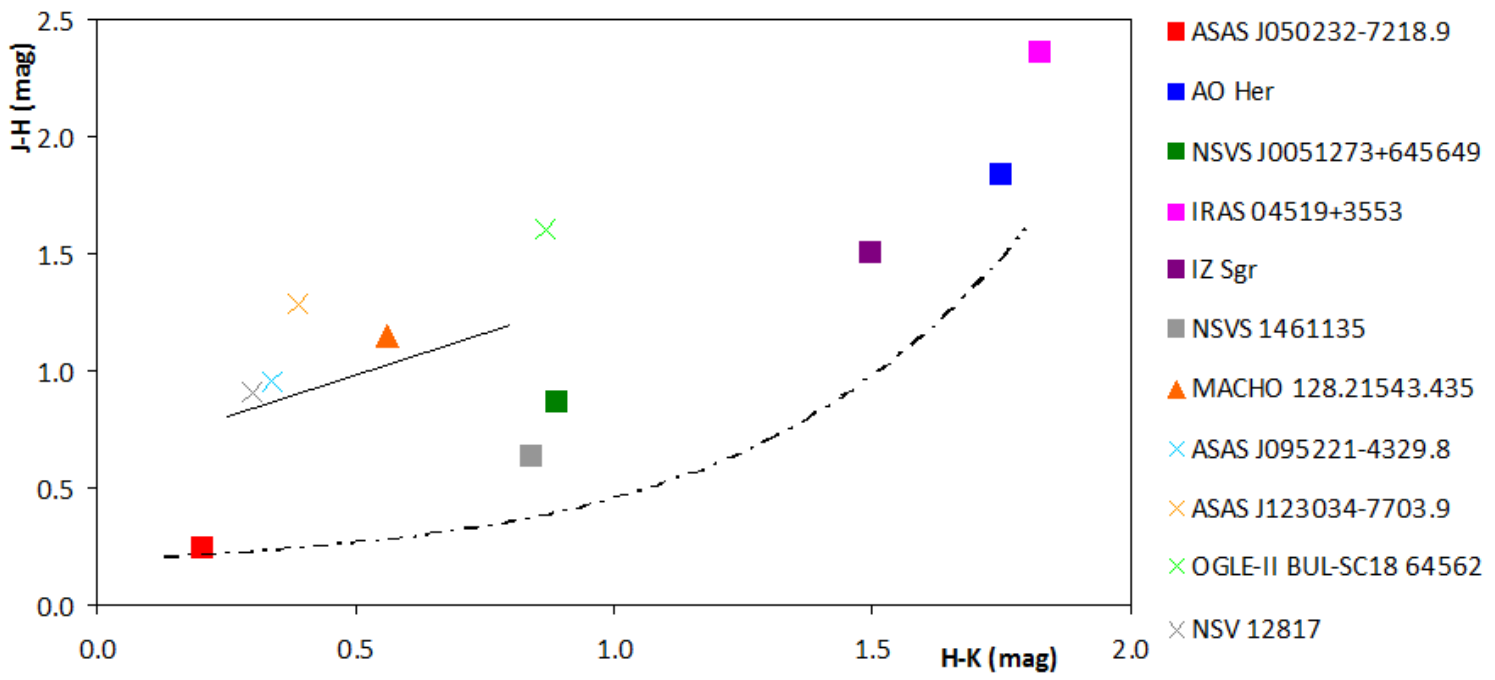

Figure 1. (J–H) vs. (H–K) diagram for all stars presented in this paper, as indicated in the legend on the right side. In order to facilitate discrimination, RCB stars and RCB candidates are denoted by squares, the proposed DYPer variable by a triangle and the non-RCB faders of section 3 by crosses. Data were drawn from the 2MASS catalogue. The solid line illustrates the colors of SMC carbon stars (Westerlund *et al.* 1991). The dashed line indicates the loci of a combination of two blackbodies, representing the photosphere of the star (~5500 K) and the dust shell (~900 K), as devised by Feast (1997).

Our findings are in very good agreement with the results of other researchers (see in particular Figure 9 of Alcock *et al.* (2001) and Figure 7 of Tisserand *et al.* (2004)), which provides additional support that the proposed classifications for the stars of the present sample, which will be enlarged on in the following sections, are valid.

### 2.2.2 (W2–W3) vs. (W3–W4) diagram

Tisserand (2012) employed mid-infrared color-color cuts based on WISE photometry to identify new RCB candidates. WISE surveyed the whole sky in the four infrared bands W1, W2, W3, and W4, which are centered at 3.4, 4.6, 12, and 22 µm, respectively (Wright *et al.* 2010). Following his approach, we have plotted a (W2–W3) vs. (W3–W4) diagram for all stars presented in this paper (Figure 2). The dashed line indicates selection cut (1) of Tisserand (2012) which effectively identifies objects with a clear shell signal (see in particular Figure 6). Because of the known bias for bright objects in the WISE W2-band, which leads to the overestimation of the brightness by up to one magnitude (see, for example, Tisserand (2012)), we have corrected the corresponding W2 magnitudes for IRAS 04519+3553 and ASAS J123034-7703.9 (Tisserand 2014; private communication). Unfortunately, WISE contamination flags indicate an uncertain value in the W2-band for ASAS J123034-7703.9.

Most RCB stars and candidates presented in this paper are well inside the denoted region on the diagram, with the exception of NSVS 1461135. The latter object's position is reminiscent of the position of MV Sgr in the respective diagram of Tisserand (2012) which shows the star well outside the expected location of RCB variables. According to the aforementioned author, MV Sgr and DY Cen—the other obvious outlier in this work—do not resemble the majority of RCB stars in that they are hot (>12000 K) and surrounded by multiple shells. It would be highly interesting to investigate if this also holds true for NSVS 1461135; however, this is beyond the scope of the present paper.

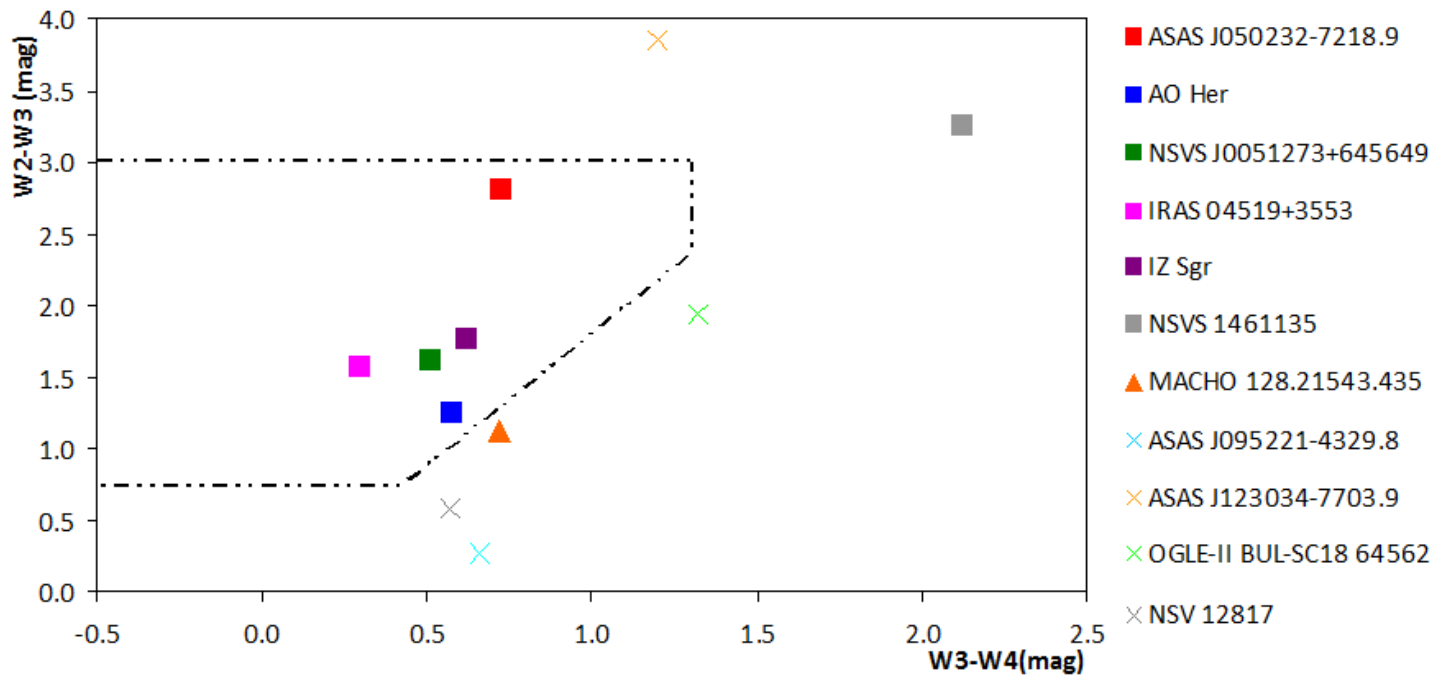

Figure 2. (W2–W3) vs. (W3–W4) diagram for all stars presented in this paper, as indicated in the legend on the right side. In order to facilitate discrimination, RCB stars and RCB candidates are denoted by squares, the proposed DYPer variable by a triangle, and the non-RCB faders of section 3 by crosses. Data were drawn from the WISE catalogue. The dashed line indicates selection cut (1) of Tisserand (2012).

## 2.3. New RCB stars

### 2.3.1. ASAS J050232-7218.9 (GSC 09169-00810)

ASAS J050232-7218.9 (GSC 09169-00810)—situated in the Large Magellanic Cloud (LMC)—is a star from the "Catalogue enriched with R CrB stars" (Tisserand 2012) that we confirm as an RCB star. The light curves of the star are shown in Figures 3 and 4.

During the first ~500 days of ASAS-3 coverage, the star's light curve is characterized by short-term variability and mean magnitude shifts between 13.6 and 14.6 magnitude (V). The short-term light changes suggest pulsational variability on a time scale of 43.1 days, although the period is rather ill-defined due to the star's brightness lying near the limiting magnitude of the survey (~14.5 magnitude (V)).

After HJD 2453000, the ASAS-3 system failed to record the star for about 1,300 days, suggesting that the object might have faded out of the survey's range during the indicated timespan. This interpretation is supported by sporadic measurements after HJD 2454438 which show the star below 15 magnitude (V). Recent data from the AAVSO Photometric All-Sky Survey (APASS; Henden *et al.* 2012) show the star at 15.65 magnitude (V) at HJD 2455557.

The reality of these isolated faint datapoints is proven by measurements from the fourth phase of the OGLE project (OGLE-IV; Udalski *et al.* 2008) which show the star recovering slowly from a low state at around 2455400 (Figure 4). During the rest of OGLE-IV coverage, which extends up to the present, the light curve is characterized by changes in mean magnitude and semiregular pulsations with a rather unstable period of ~35 days. Additional proof that ASAS J050232-7218.9 is capable of deep fadings comes from the Magellanic Clouds Photometric Survey (Zaritsky *et al.* 2004), in which the object is listed with a V magnitude of 18.1. This measurement was taken at HJD 2451517 (Zaritsky 2014).

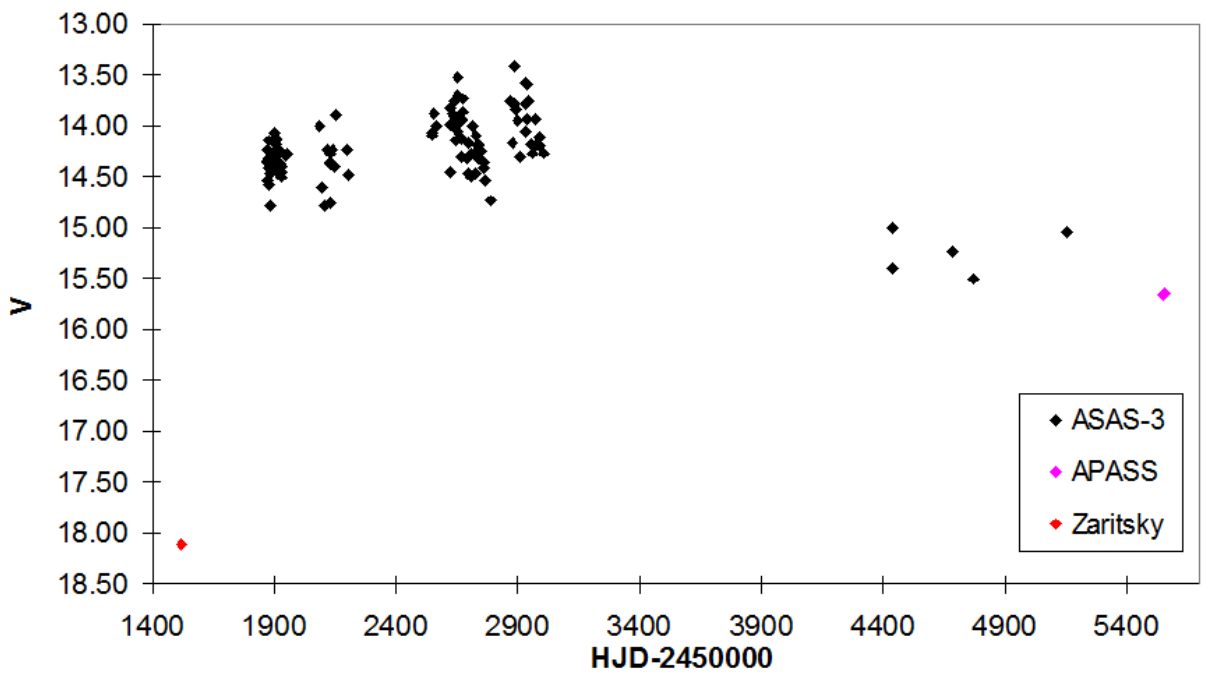

Figure 3. Light curve of ASAS J050232–7218.9, based on ASAS-3 data, APASS data, and data from the Magellanic Clouds Photometric Survey (Zaritsky *et al.* 2004).

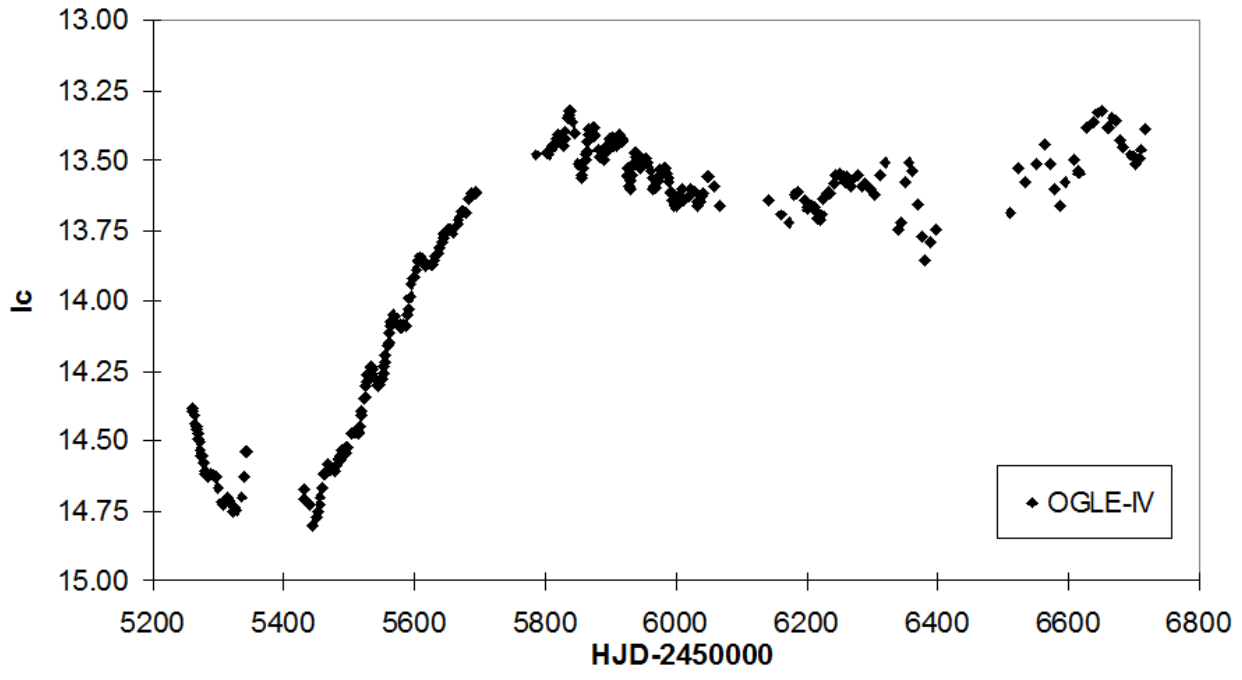

Figure 4. Light curve of ASAS J050232–7218.9, based on OGLE-IV data.

Color measurements from 2MASS (J–K = 0.46) and APASS (B–V = 1.34) are in agreement with values found for other RCB stars, which also holds true for the star's position in the near- and mid-infrared two-color diagrams (section 2.2). Finally, and most importantly, the star has been observed spectroscopically in the past. It had been classified as an R-type carbon star by Hartwick and Cowley (1988) and was found to be a hydrogen-deficient carbon star (classified as C2,2Hd) in Richer *et al.* (1979). The other three hydrogen-deficient stars published in the latter paper were subsequently confirmed as RCB stars (EROS2-LMC-RCB-2, EROS2-LMC-RCB-3, and EROS2-LMC-RCB-5;

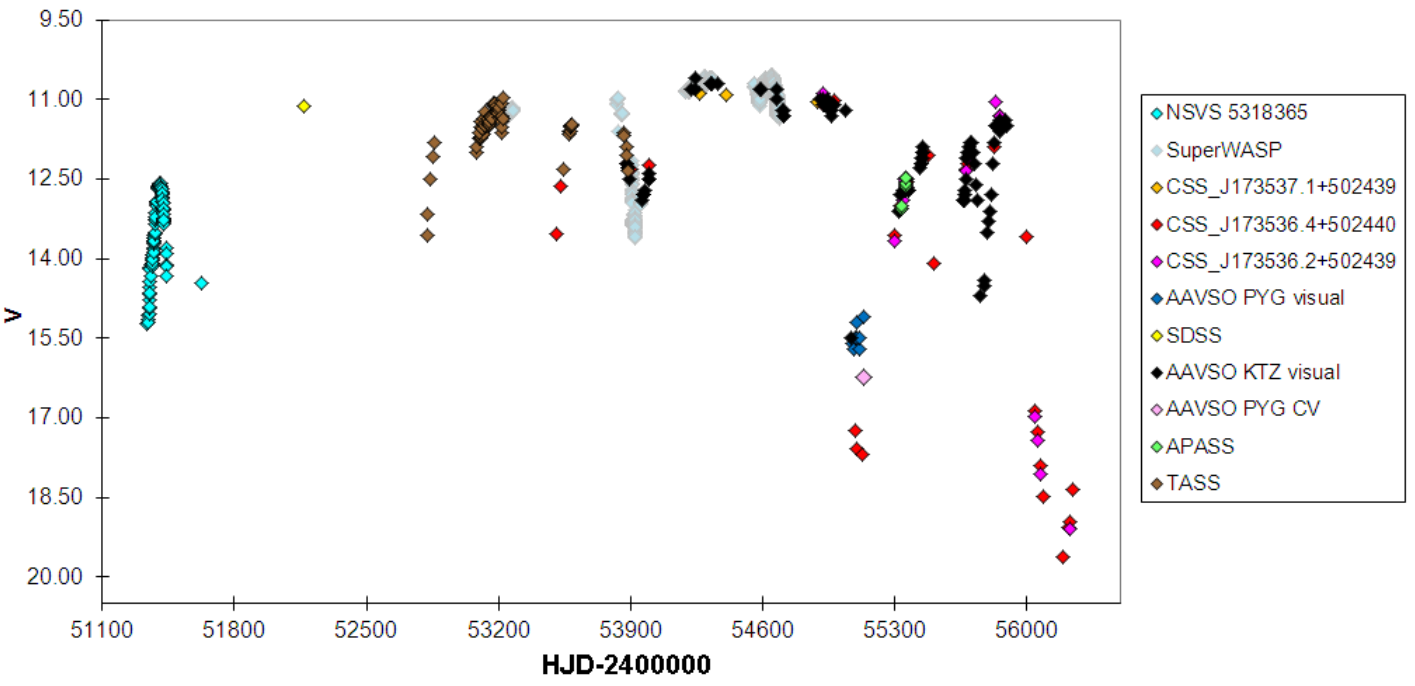

Figure 5. Light curve of AO Her, based on data from various sky surveys, as indicated in the legend.

Tisserand *et al.* 2009). We now confirm the remaining one, ASAS J050232-7218.9, as an RCB variable on grounds of its photometric behavior.

2.3.2. AO Her (2MASS J17353628+5024398)

AO Her (2MASS J17353628+5024398) was discovered in 1924 by Woods at Harvard Observatory (Woods 1924). In an early study of the star, Böhme (1937) comments on AO Her's flat and broad maxima and characterizes its periodicity as semiregular at best. He lists five times of maxima which have been deduced from 80 observations spread among about 3,000 days (between JD 2425688 and JD 2428717), emphasizing their uncertainty because of the star's light curve properties. In the specified period, AO Her's photographic brightness lay between 11.4 magnitude (p) on JD 2427645 and <13.0 magnitude (p) outside of maximum observations. The star has been poorly studied during the following decades but it has been included into the RCB-enriched catalogue of Tisserand (2012). A combination of all available data gives proof of several significant fading events during the covered timespan (Figure 5).

Data from the NSVS show AO Her rising from 15.4 magnitude (ROTSE-I) to about 12.5 magnitude (ROTSE-I) at around HJD 2451400, after which it drops back to around 14.4 magnitude (ROTSE-I). ROTSE-I magnitudes from the NSVS are unfiltered and calibrated with V magnitudes (see Akerlof *et al.* 2000) but are probably much brighter than the visual scale in this case, especially as AO Her is a red object. However, we did not shift NSVS data as there are no contemporaneous V measurements of AO Her to allow a reliable calibration. The NSVS light curve suggests the typical behavior of a red variable, supporting the classification given in the original paper.

There follows a large gap in available data for AO Her between HJD 2451600 and HJD 2452800, with only a single measurement from the Sloan Digital Sky Survey (SDSS; Adelman-McCarthy *et al.* 2011) which shows the star at around 11.0 magnitude (V). The situation improves significantly with the onset of The Amateur Sky Survey (TASS; Richmond 2007) and the SuperWASP project (SWASP; Butters *et al.* 2010) and—at around HJD 2453900—regular visual observations made by AAVSO observers (AAVSO 2013) that have been calibrated with the APASS V scale.

In addition to several minor drops of about 2 to 3 magnitudes (V), two major declines are recorded. The first one took place at around HJD 2455150, when AO Her was as faint as 17.7 magnitude (V). At the end of the covered timespan, data from the Catalina Sky Survey (CSS; Drake *et al.* 2009)—shifted ~1 magnitude to match the V scale from APASS and AAVSO visual data—show the star at its faintest, recovering from below 19.6 magnitude (V) on HJD 2456195, indicating that AO Her's amplitude exceeds 9 magnitudes (V).

In summary, it can be stated that AO Her exhibits light curve properties typical of RCB stars, in particular the sudden and dramatic fadings characteristic of this kind of variable stars. Additionally, color measurements from 2MASS (J–K = 3.59) and APASS (B–V = 1.86) are indicative of infrared excess, and the position of AO Her in the two-color diagrams is characteristic of an RCB star. The classification of AO Her as a bona fide RCB star seems, therefore, justified.

At the time of this writing, AO Her has been confirmed spectroscopically as an RCB variable by Tisserand *et al.* (2013a) with the FLOYDS spectrograph on the 2-m LCOGT/Faulkes Telescope North. Their results, along with data from a planned monitoring campaign, will be published in a future paper.

## 2.4. RCB / DYPer candidates

### 2.4.1. NSVS J0051273+645649 (GSC 04025-00779)

The variability of NSVS J0051273+645649 (GSC 04025-00779) was discovered during a search for red variables in the Northern Sky Variability Survey (NSVS) by Woźniak *et al.* (2004a). We have identified the star as a likely RCB variable by investigation of candidates from the RCB-enriched catalogue of Tisserand (2012). The light curve of the star (shifted to the APASS V zero point) is shown in Figure 6.

The light curve of NSVS J0051273+645649 is characterized by semiregular pulsations with a predominant period of 29.8 days (Figure 7) and a significant fading event at around HJD 2451500. The star faded by about 1.5 magnitudes (ROTSE-I), showing signs of a slow recovery during the rest of NSVS coverage, which is typical of RCB stars.

Color measurements from 2MASS (J–K = 1.76) and APASS (B–V = 1.48) give evidence of infrared excess. In addition, the star's B–V index is too blue to qualify it as a DYPer variable. The proposed classification is also supported by the star's position in the near- and mid-infrared two-color diagrams (section 2.2). We therefore conclude that NSVS J0051273+645649 is a likely RCB candidate worthy of follow-up investigations.

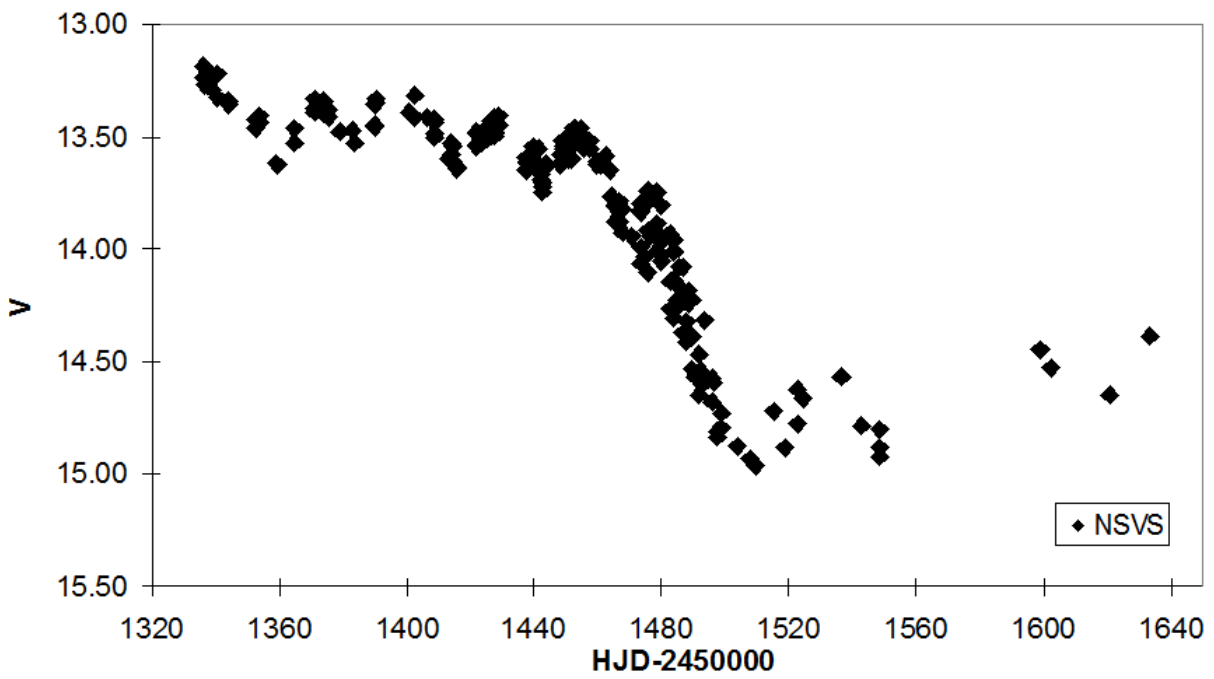

Figure 6. Light curve of NSVS J0051273+645649, based on NSVS data; ROTSE-I magnitudes from NSVS were shifted to the V scale.

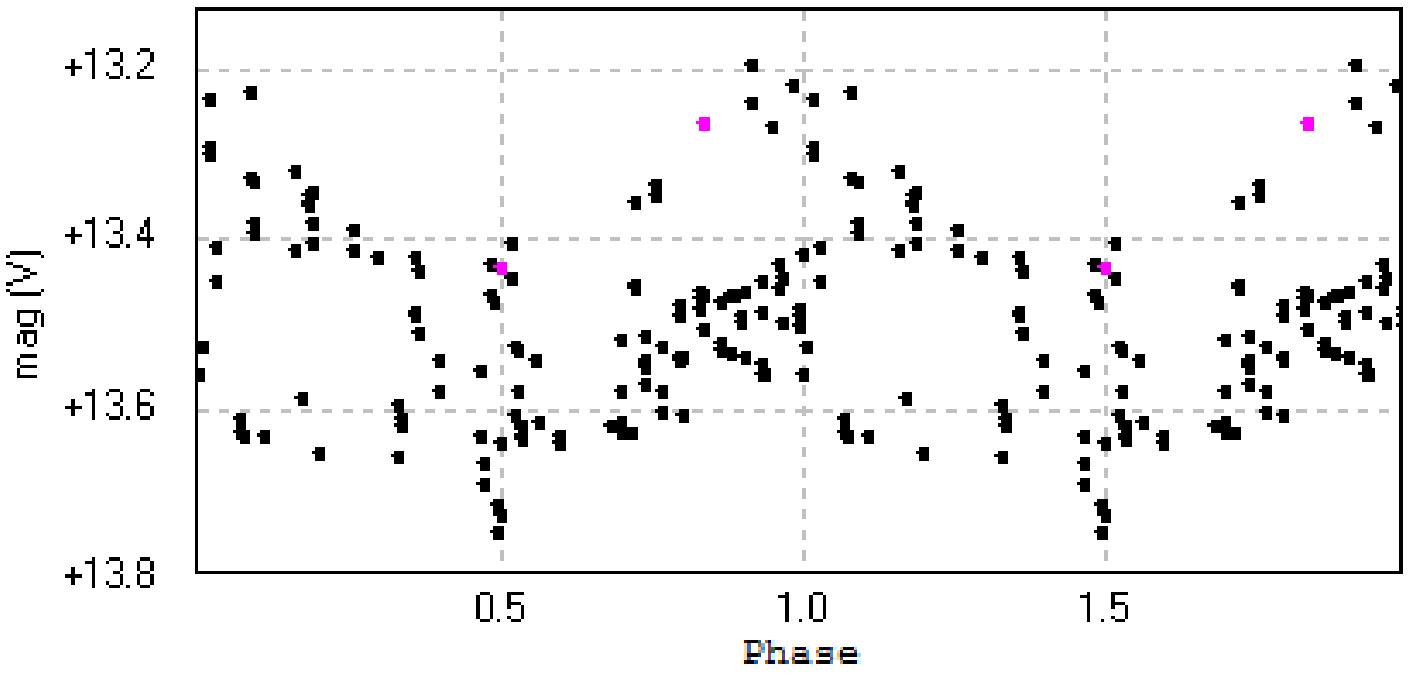

Figure 7. Semiregular pulsations of NSVS J0051273+645649; the phase plot has been based on NSVS data (2451335 < HJD < 2451464; black points) and APASS data (pink points) and folded with a period of P = 29.8 days. ROTSE-I magnitudes from NSVS were shifted to the V scale.

### 2.4.2. IRAS 04519+3553 (2MASS J04552045+3558079)

IRAS 04519+3553 (2MASS J04552045+3558079) is another large amplitude variable from the RCB-enriched catalogue of Tisserand (2012) which has exhibited conspicuous fading events in the past. Timeseries photometry for this object is available from the CSS; unfortunately, the light curve is sparsely sampled, consisting of 160 datapoints only, and frequently interrupted by observational gaps. Thus, the exact shape of the light curve is open to conjecture during some parts of the coverage (Figure 8).

Despite the above-mentioned difficulties, there are indications of several fadings during the covered timespan. Four closely spaced measurements at the beginning of CSS coverage show the star at ~12.5 magnitude (CV). However, following a gap of ~200 days, the star is found rising from below 14.1 magnitude (CV), which suggests that a drop in brightness has occurred during the observational gap. A more conspicuous fading event is centered around

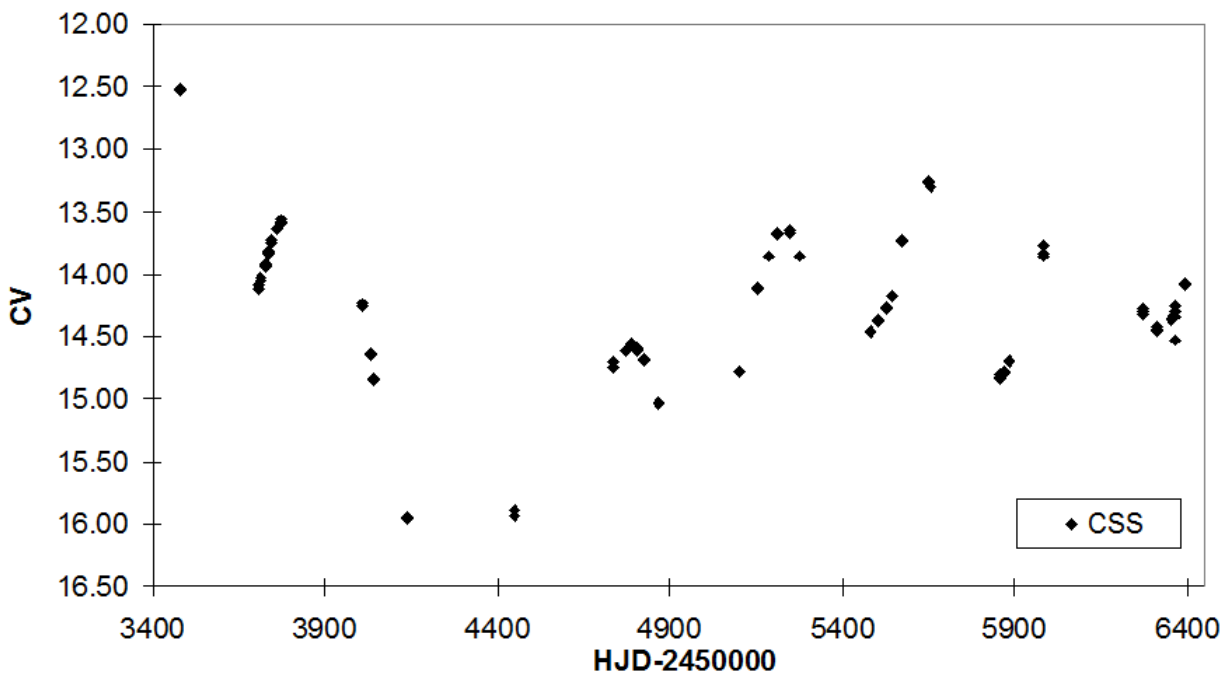

Figure 8. Light curve of IRAS 04519+3553, based on CSS data.

HJD 2454300, with the star dropping sharply from 14.25 magnitude (CV) to about 16.0 magnitude (CV). Again, an observational gap in the data leaves the amplitude and the exact shape of the minimum open to conjecture. Another possible fading event might have taken place around HJD 2455000.

During the rest of CSS coverage, the light curve is reminiscent of large amplitude pulsations, which is unusual for an RCB variable. An analysis of the available data yields a period of P~438 days, a value commonly observed, for example, in carbon-rich semiregular variables that are also sometimes prone to fading events. However, similar pulsations have been reported in RCB stars—for example, EROS2-CG-RCB-12 (Tisserand *et al.* 2008) and EROS2-LMC-RCB-8 (Tisserand *et al.* 2009)—albeit with shorter periods. It remains up to debate whether these light curve features are actually pulsations or connected to small-scale fading events (see Tisserand *et al.* 2009).

A classification as a classic carbon star would also agree with the observed J–K index of 4.18 (2MASS). However, the star's position in the near- and mid-infrared two-color diagrams of section 2.2 is in agreement with a classification as an RCB variable and provides evidence against the former assumption. The star is situated in the "all shell" region of the (J–H) vs. (H–K) diagram (Figure 1). Together with the observed J–K index of 4.18, which is indicative of infrared excess, this implies that 2MASS observations (epoch: HJD 2451093.8724) were made during an obscuration event when the object was deeply enshrouded in dust.

Measurements from the GSC2.3 catalogue and CMC14 (Evans *et al.* 2002; transformed to V) indicate a V magnitude of 18.15 magnitude (epoch: HJD 2447860.5) and 18.3 magnitude (epoch: HJD 2452649.5), respectively. However, it is not possible to tell if the star was in a faint state at the time of the measurements or if it is intrinsically faint at visual wavelengths, which would imply that IRAS 04519+3553 is a very red object even at maximum. This would account for the large difference between the object's brightness in

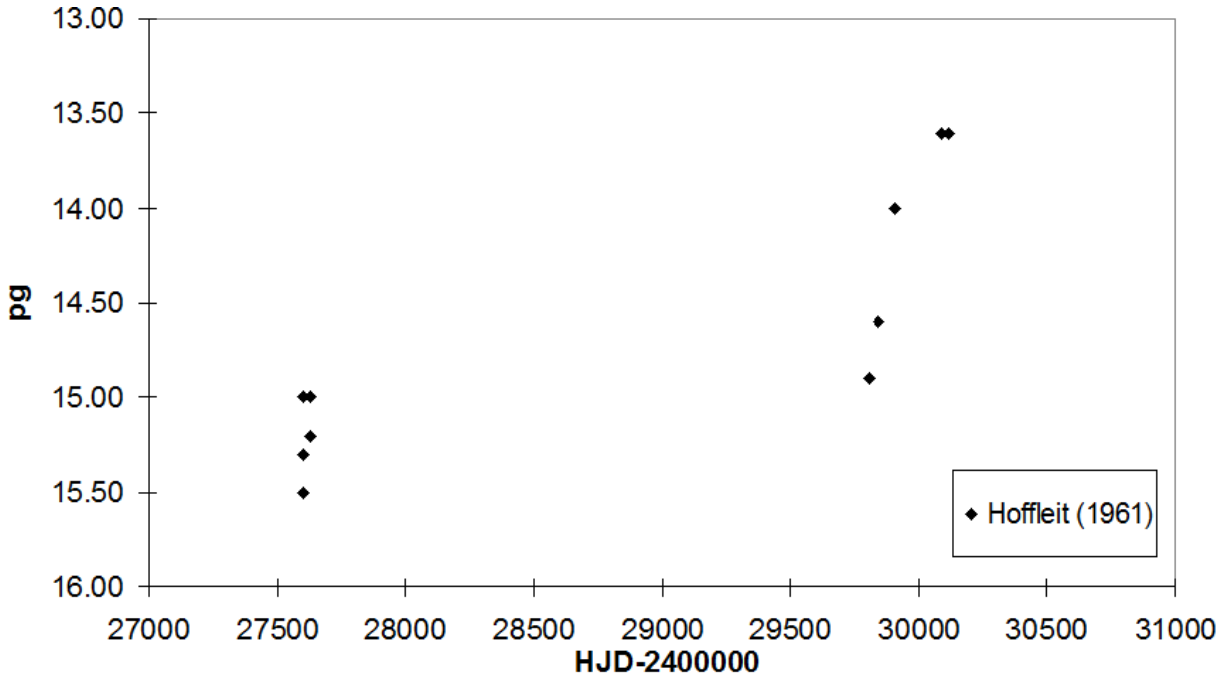

Figure 9. Light curve of IZ Sgr, based on data from Hoffleit (1961).

V and Catalina unfiltered photometry, which is calibrated against V magnitudes but strongly dependent on source color, being more sensitive to the red portion of the spectrum.

Taking into account the above-mentioned evidence, we are unable to arrive at a conclusive classification for IRAS 04519+3553, which we consider a likely RCB candidate. Long-term photometric monitoring and spectroscopic studies are encouraged.

2.4.3. IZ Sgr (GSC 06279-00870)

The star was discovered as HV 4148 by Woods (1928) and later designated as IZ Sgr in the *General Catalogue of Variable Stars* (GCVS; Samus *et al.*, 2007–2013). It was included in an investigation of 45 variable stars by Hoffleit (1961), who commented on the object's invisibility on the majority of the available plate material; positive observations of IZ Sgr were possible on only a dozen of several hundred plates reaching to below 15th magnitude (pg) that were available to Hoffleit (Figure 9).

An investigation of ASAS-3 data presents a similar picture. Except for two bright phases—one Mira-like hump at around HJD 2452100 and one rather broad maximum from about HJD 2453050 to HJD 2453450—the star remained constantly below the survey's detection limit (Figure 10). Therefore, it is evident that IZ Sgr is capable of unpredictable steep rises and drops in magnitude; additionally, a classification as a Mira variable can be excluded. The light curve also shows indications of pulsational variability, in particular during the rise to maximum light near HJD 2453100.

Measurements from various catalogues indicate a large range for IZ Sgr (for example: B = 13.3 (YB6)–19.1 (USNO-B1.0); R = >12.9 (USNO-A2.0)–<17.6 (USNO-B1.0)). Furthermore, there is no entry for IZ Sgr in CMC14 or SPM4.0 (Girard *et al.* 2011), while the 17.8 magnitude (V) star SPM4.0 6551155866, which lies only 7" away, is recorded in both catalogues. This suggests that IZ

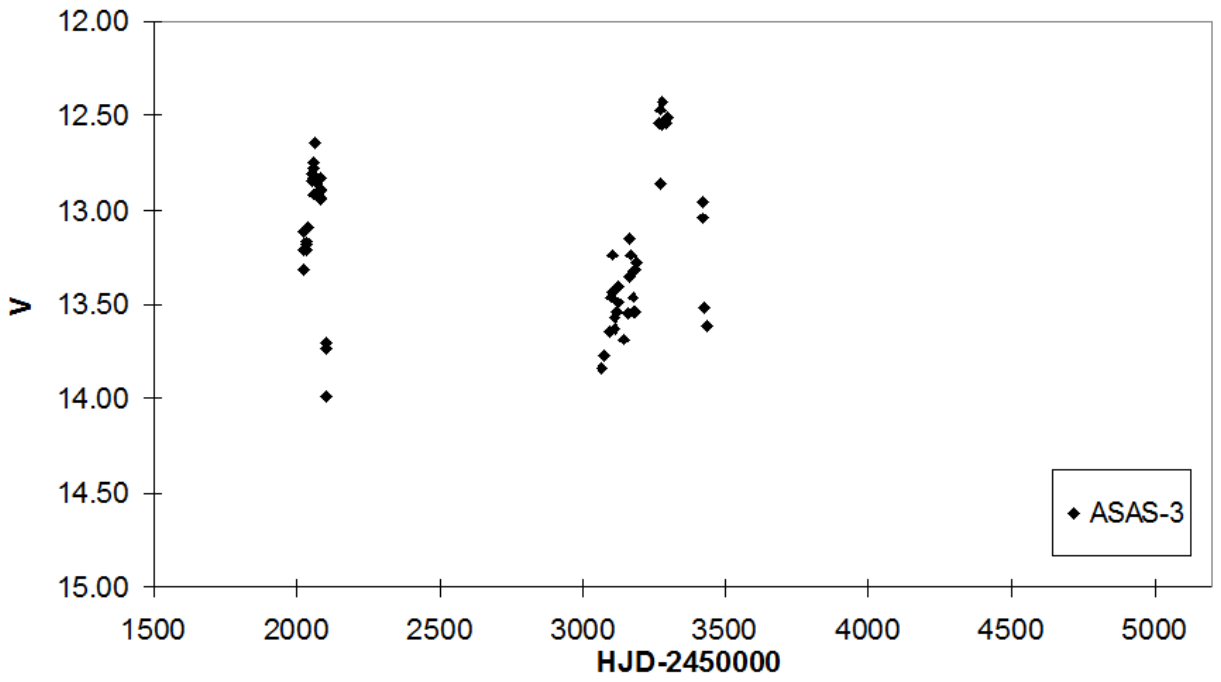

Figure 10. Light curve of IZ Sgr, based on ASAS-3 data.

Sgr might even drop to below 17.8 magnitude (V). The observed amplitude and behavior of IZ Sgr are rather extreme and do not resemble the variations seen in most irregular L-type variables.

Additionally, Tisserand *et al.* (2013b) reported that the star was too faint for spectroscopic follow-up. The star was about V = 17 magnitude during June and July 2012 (Tisserand 2014).

Although its listed spectral type of M6 (Houk 1967) is not in agreement with a classification as an RCB variable, the star's position in the near- and mid-infrared two color diagrams is consistent with its classification as an RCB variable (section 2.2). Considering the possibility that the assigned spectral type might be in error, we strongly encourage further photometric and spectroscopic investigations to gain an insight into the nature of IZ Sgr, which we consider a strong RCB candidate.

2.4.4. NSVS 1461135 (GSC 04282-00656)

The variability of NSVS 1461135 (GSC 04282-00656), which is situated in the field of the open cluster [KPR2005] 125 (Zejda *et al.* 2012), was discovered during a search for variable Asymptotic Giant Branch (AGB) stars in the NSVS database by Usatov and Nosulchik (2008), who classified it as a candidate Mira variable. The star's light curve is characterized by rather irregular pulsations and a significant fading at around HJD 2451375 (Figure 11). The fast decline and slow recovery from this event is reminiscent of the photometric behavior of RCB stars, although the amplitude is small (~1.2 magnitude (ROTSE-I)).

The (unfiltered) ROTSE-I magnitudes of NSVS 1461135 are similar to the V magnitudes that have been gleaned from various catalogues (for example: 13.56 magnitude (UCAC3) and 13.10 magnitude (TASS)), which would be highly unusual for a red object. Furthermore, from two APASS observations at 12.93 magnitude (V) and 12.95 magnitude (V), we derive a color index of B–V = 0.63, which also suggests a yellow star and, in combination with the object's J–K index of 1.47 (2MASS), is indicative of infrared excess. This is in agreement with an RCB classification and strong evidence against NSVS 1461135 being a DYPer variable, which is also supported by its position in the near-infrared two-color diagram (section 2.2.1). The star's deviant position in the mid-infrared two-color diagram does not necessarily disqualify NSVS 1461135 as an RCB candidate, as similar results have been reported in the literature (section 2.2.2).

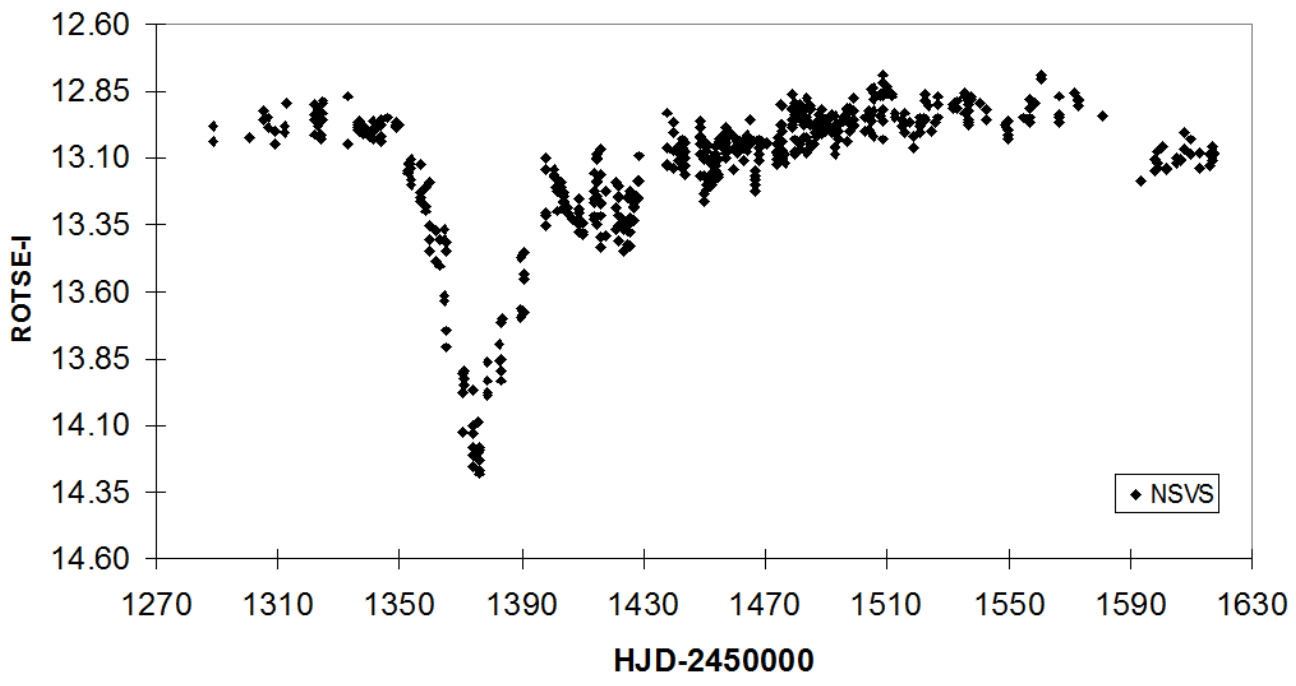

Figure 11. Light curve of NSVS 1461135, based on NSVS data.

However, the observed range is not large; the faintest recorded magnitude is 14.2 (V), which we derived from SDSS photometry. A classification as a different type of variable—notably a long-period eclipsing binary star like the symbiotic systems V5569 Sgr and V1413 Aql—is not excluded. Further photometric and spectroscopic analyses are advised to arrive at a conclusive classification.

2.4.5. MACHO 128.21543.435 (2MASS J18063154-2834301)

During a search for new Mira variables in the MACHO Galactic Bulge fields (see, for example, Hümmerich and Bernhard 2012), MACHO 128.21543.435 (2MASS J18063154-2834301) was found to exhibit significant fading events during the covered timespan. Additional data were procured from the EROS-2 project (Renault *et al.* 1998; Tisserand and Marquette 2014). The light curve of the star is shown in Figure 12. The transformation of MACHO instrumental magnitudes to the Kron-Cousins system was done by using Equation (2) of Alcock *et al.* (1999). EROS-2 data have been transformed to Johnson V and Cousins $I_c$ using Equation (4) of Tisserand *et al.* (2007).

The first obscuration event took place around HJD 2449200, with the star dropping from about 15.0 magnitude ($R_c$) to 17.15 magnitude ($R_c$). A second decline occurred around HJD 2450800; the minimum magnitude is open to conjecture because of an observational gap in the data. Although both declines are only partially covered by MACHO, they suggest the symmetric fading events and sharp minima of a DYPer star. Another fading event was recorded by the EROS-2 project and took place at around HJD 2452300. Again, the amplitude of the fading can only be estimated due to an observational gap in the data; it exceeds at least 1.5 magnitudes in both V and $I_c$, though.

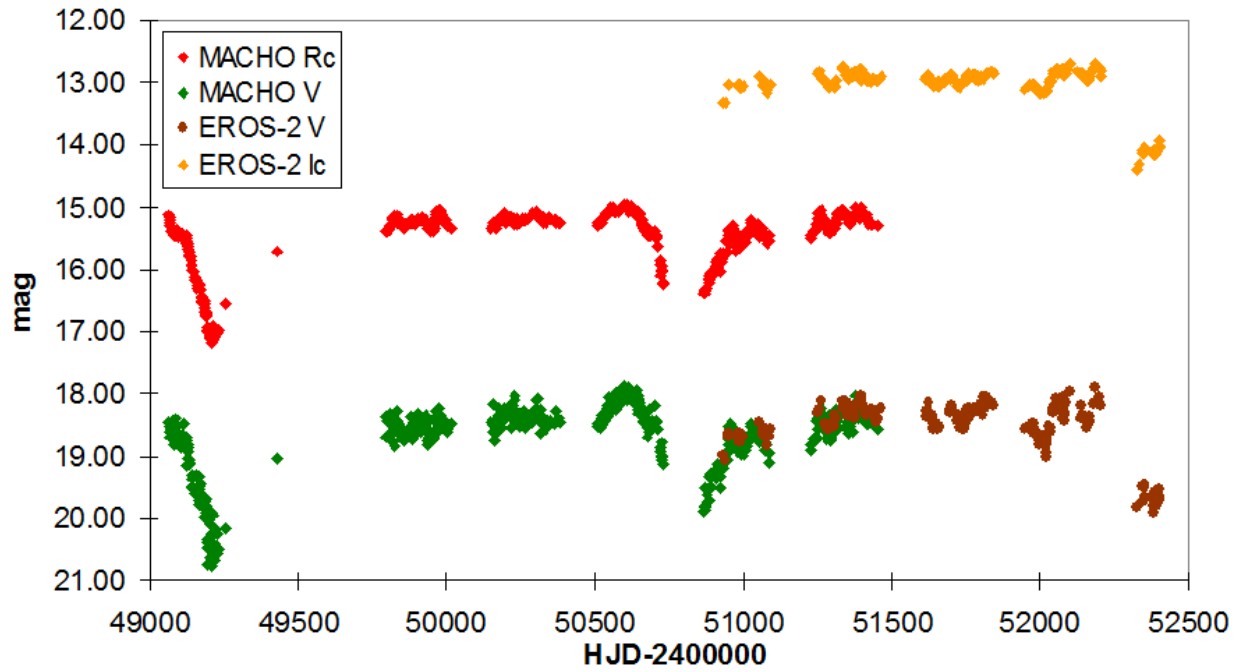

Figure 12. $VR_cI_c$ light curve of MACHO 128.21543.435, based on data from the MACHO and EROS-2 projects. Obvious outliers have been removed by visual inspection.

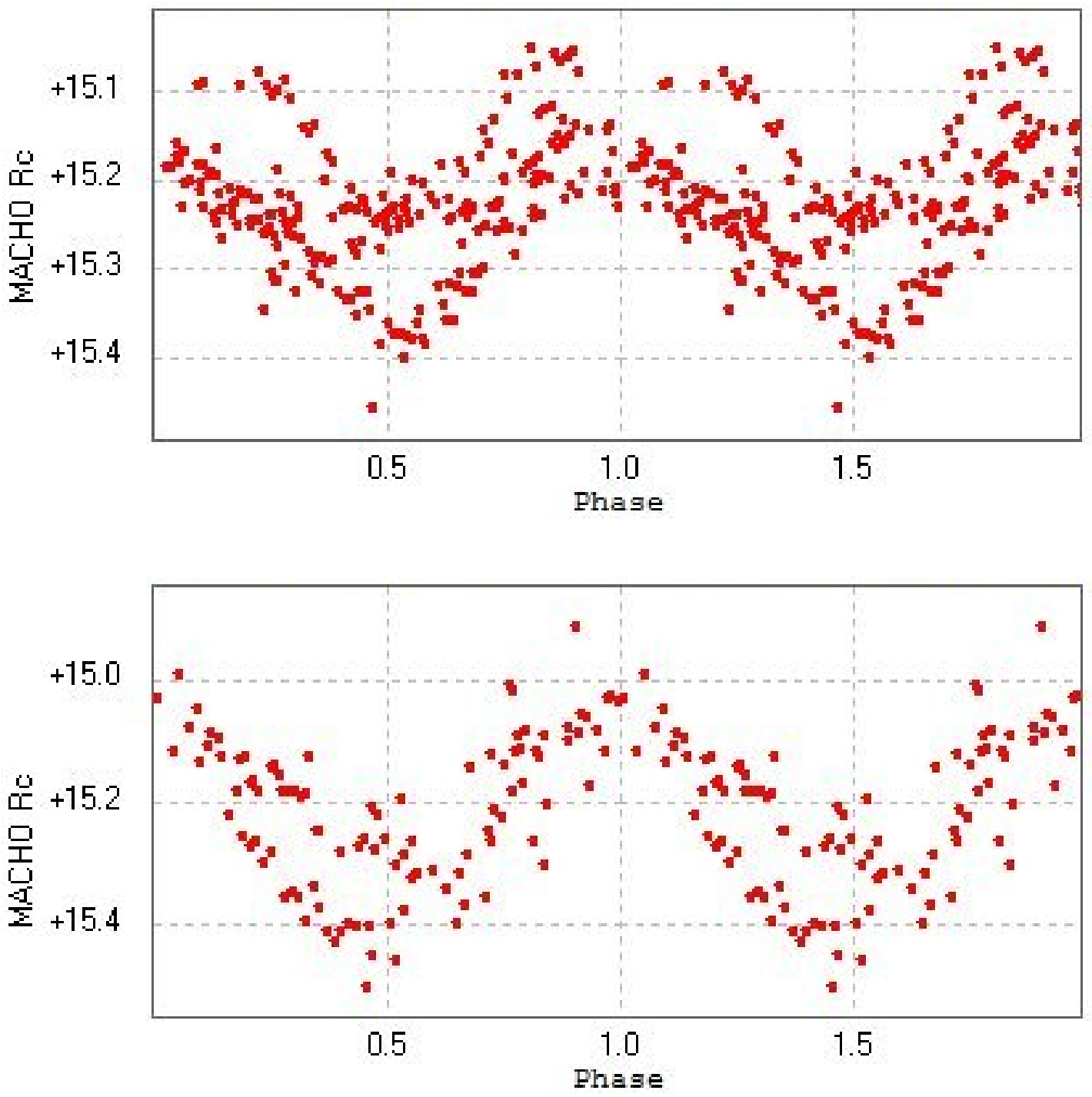

Figure 13. Semiregular pulsations of MACHO 128.21543.435; the phase plots have been based on MACHO $R_c$ data. Phase plot 1 (top panel) is folded with P = 75.6 days (2449794 < HJD < 2450380), phase plot 2 (bottom panel) is folded with P = 66.4 days (2451229 < HJD < 2451454).

Outside these obscurations, the light curve is characterized by semiregular pulsations. A period of 75.6 days is predominant in the timespan from HJD 2449794 to HJD 2450380, which then changes to 66.4 days in the data from HJD 2451229 to HJD 2451454 (Figure 13).

Judging from its light curve properties (notably the symmetric fadings with rapid declines and sharp minima) and its position in the near-infrared two-color diagram (section 2.2.1), which is in agreement with that of the DYPer star candidates of, for example, Alcock *et al.* (2001) and Tisserand *et al.* (2004), we conclude that MACHO 128.21543.435 is a promising DYPer candidate. However, spectroscopy is needed for a conclusive classification.

## 3. Photometrically-related objects of interest

All objects presented in this section show photometric behavior similar to RCB stars—namely, significant, unpredictable fading events and semiregular pulsations. Considering spectra, color information, and/or light curve peculiarities, it is evident, though, that these objects are not RCB variables; in fact, some of them have been rejected as such on spectroscopic grounds in recent investigations.

Most of the stars are likely to be ordinary red giants or carbon stars undergoing obscuration events; however, some of them are not easy to assign to a type and their peculiar behavior might merit more detailed follow-up studies. The common link behind the range of observed behavior in these objects seems to be dust ejection on significant scales. It would be highly interesting to investigate what—if anything—differentiates these objects from standard, non-fading semiregular red giant stars. It is possible that long term photometric coverage of red giants would considerably increase the number of these stars, whose behavior might be due to short-lived evolutionary processes.

### 3.1. Notes on individual stars

#### 3.1.1. ASAS J123034-7703.9 (GSC 09416-00380)

ASAS J123034-7703.9 (GSC 09416-00380) is listed in the *ASAS Catalog of Variable Stars* (ACVS; Pojmański *et al.* 2005) as a miscellaneous variable star (type "MISC") and was proposed as an RCB candidate by Hümmerich (2011). It exhibits a significant obscuration event in its light curve, dropping about two full magnitudes (V) at around HJD 2453000 and remaining in this faint state for the rest of ASAS-3 coverage (Figure 14).

The star's light curve is further characterized by large-amplitude pulsations with a period of about 273 days that are strongly affected by the fading event, during which the pulsational amplitude shrinks to about 0.2 magnitude (V).

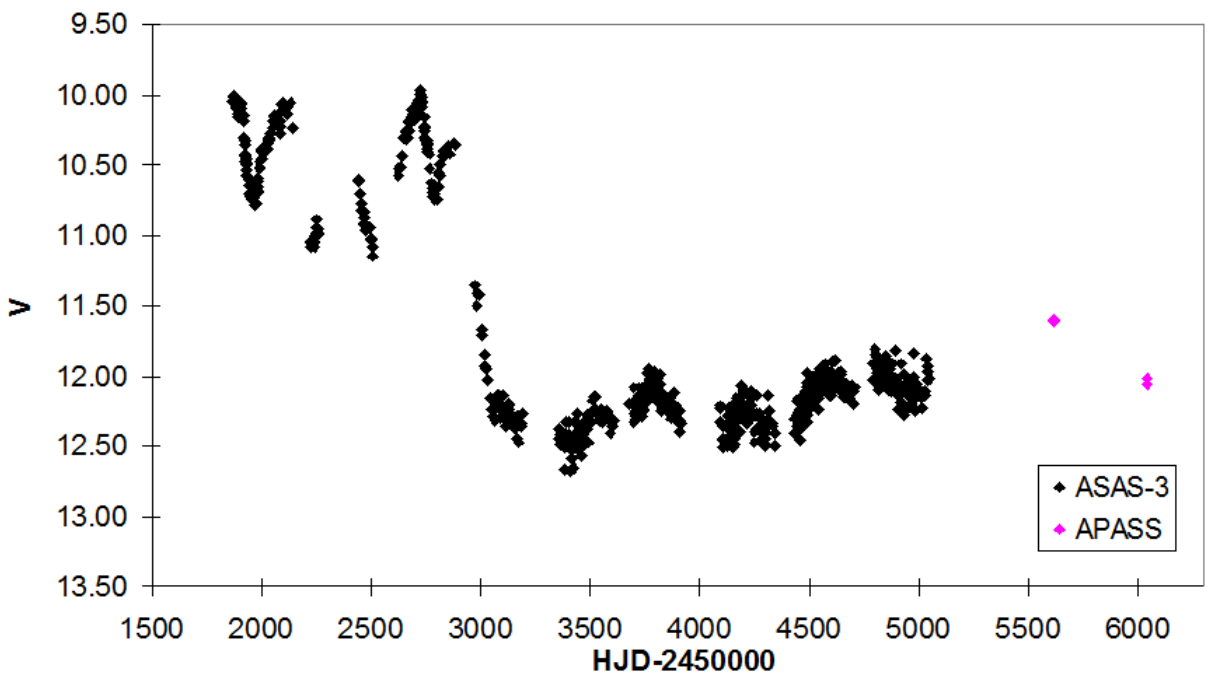

Figure 14. Light curve of ASAS J123034–7703.9, based on ASAS-3 and APASS data. Obvious outliers have been removed by visual inspection.

Color measurements from 2MASS (J–K = 1.67) and APASS (B–V = 2.26) indicate that ASAS J123034-7703.9 is a red object which is also very bright in the near infrared (J = 3.38; H = 2.10; K = 1.71 (2MASS)). In the near-infrared two-color diagram, ASAS J123034-7703.9 is positioned near the loci of classical carbon stars (section 2.2). The star's position in the (W2–W3) vs. (W3–W4) diagram hints at the existence of a cool circumstellar dust shell (Figure 2), which is in agreement with the observed prolonged obscuration event.

Amplitude and period of the pulsations as well as color information are typical of a red giant star. Additional information comes from an IRAS Low Resolution Spectrum (Joint IRAS Science Working Group 1987) that shows Si emission at 9.7 µm (Figure 15), which is known to originate from the circumstellar environments of oxygen-rich AGB stars (for example, Kwok *et al.* 1997). Thus, a classification as RCB or DYPer variable seems highly unlikely; in fact, Miller *et al.* (2012) identified ASAS J123034-7703.9 as a potential candidate but ultimately rejected it on grounds of the aforementioned IRAS spectrum and the classification in Kwok *et al.* (Miller 2012).

It seems that ASAS J123034-7703.9 is another example of a red giant undergoing a significant fading event because of an episode of dust formation, as has been shown, for example, for L2 Pup (Bedding *et al.* 2002). It is interesting to note that the amplitude of the pulsational variations of ASAS J123034-7703.9 has been affected much more strongly by the obscuration event than in the case of L2 Pup. The light curve of L2 Pup from 1980 to the present time is shown in Figure 16.

### 3.1.2. OGLE-II BUL-SC18 64562 (2MASS J18064655-2722063)

During an investigation of variable objects with high amplitudes from the OGLE Galactic Bulge fields using OGLE-II data (Udalski *et al.* 1997; Szymański 2005), a significant fading event was discovered in the light curve of OGLE-II BUL-SC18 64562 (2MASS J18064655-2722063). As shown in Figure 17, the star was measured at ~13.4 magnitude ($I_c$) before an observational gap at around HJD 2450800, during which it apparently dropped by more than 2 magnitudes ($I_c$). The object was next detected at HJD 2450860 with a brightness of around 16.0 magnitude ($I_c$). It started to rise to 14.5 magnitude ($I_c$) shortly after, at which brightness it remained during the remainder of OGLE-II coverage.

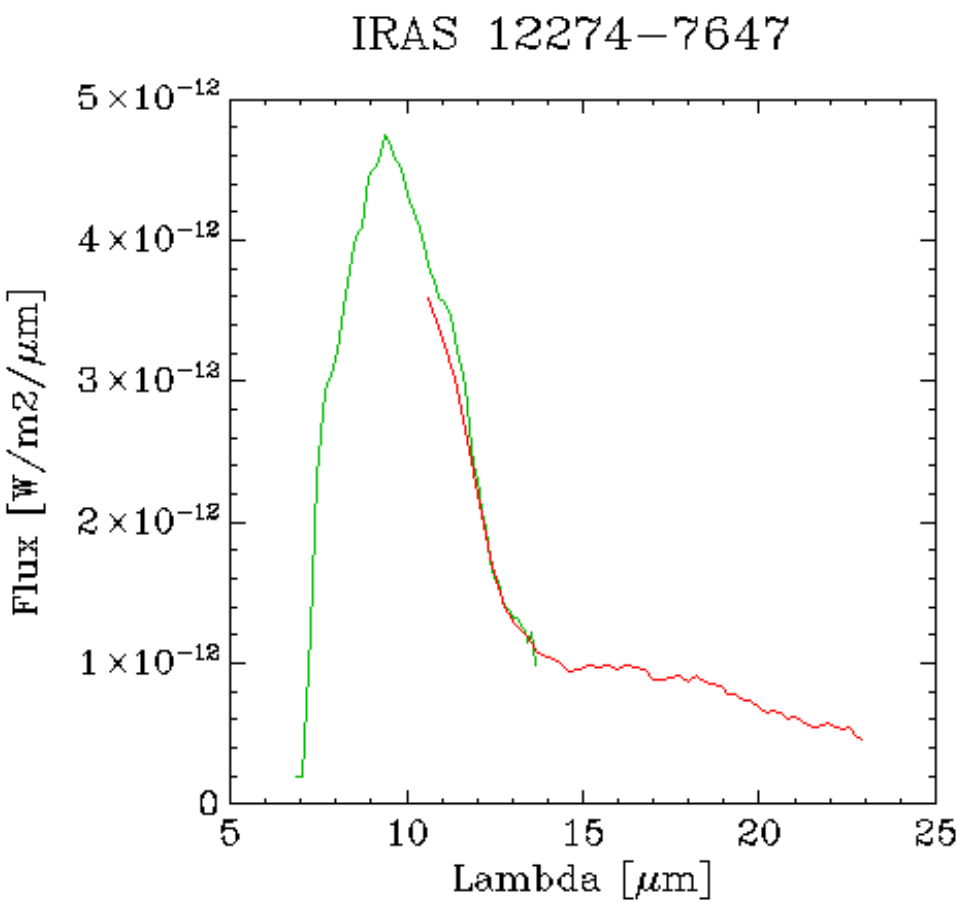

Figure 15. IRAS low resolution spectrum of ASAS J123034–7703.9 showing Si emission (Joint IRAS Science Working Group 1987).

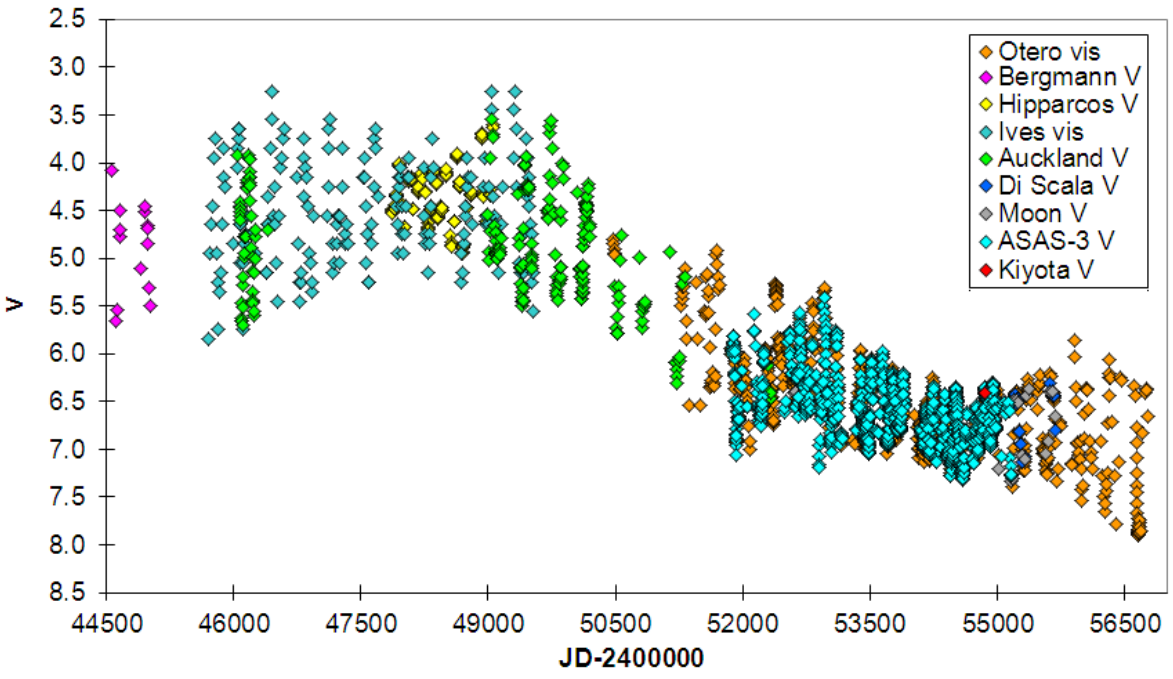

Figure 16. Light curve of L2 Pup, based on various data sources, as indicated in the legend.

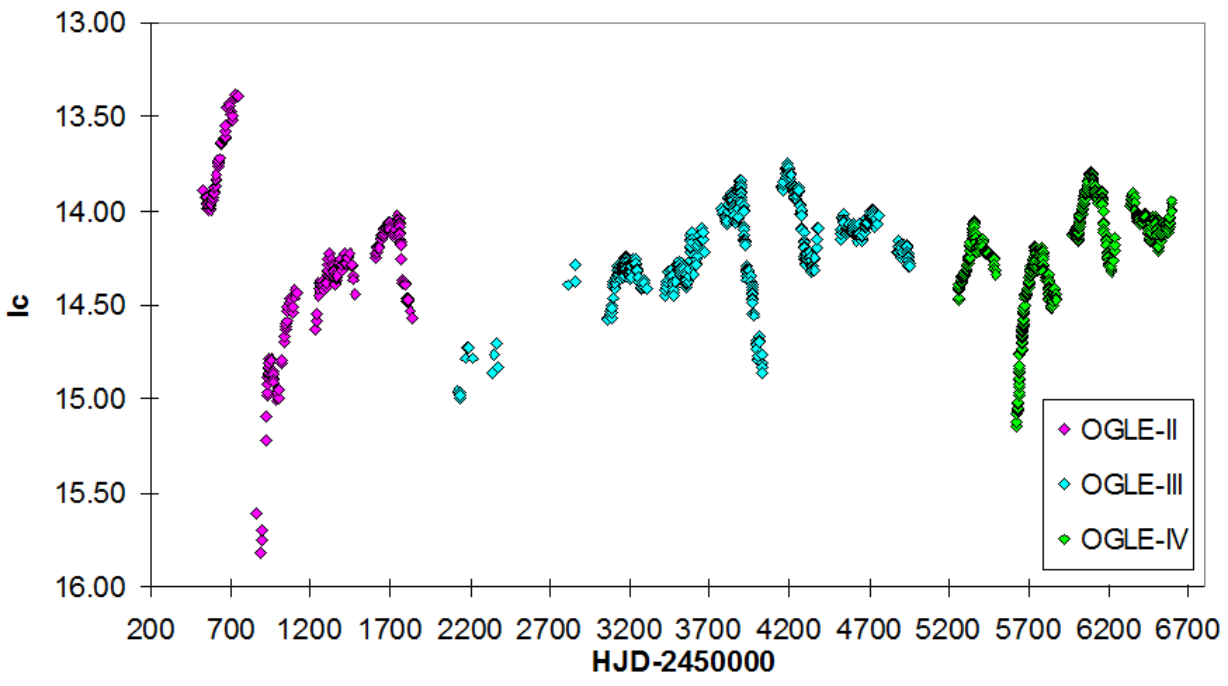

Figure 17. Light curve of OGLE-II BUL-SC18 64562, based on OGLE-II, OGLE-III, and OGLE-IV data.

Additional data were procured from the *OGLE-III Catalog of Long-Period Variables (LPVs) in the Galactic Bulge* (Soszyński *et al.* 2013), which lists the object as an OGLE small amplitude red giant (OSARG), and from OGLE-IV. In addition to a long-term mean magnitude shift, three more fading events were revealed at around HJD 2452000, HJD 2454000, and HJD 2455600, albeit of much smaller scale. The light curve is further characterized by low-amplitude pulsations of a rather irregular nature, although a period of ~60 days is recognizable during parts of the coverage, for example, around HJD 2451300, HJD 2453800 and HJD 2456500.

OGLE-II BUL-SC18 64562 is a very red object with color indices of V–I = 4.92 (Udalski *et al.* 2002) and J–K = 2.47 (2MASS). It is positioned among the classical carbon stars in the (J–H) vs. (H–K) diagram (section 2.2.1), thus joining the ranks of red variables undergoing significant obscuration events. Further photometric and spectroscopic studies are encouraged.

3.1.3. NSV 12817 (CD-50 12825) and ASAS J095221-4329.8 (CD-42 5700)

CD-50 12825 was announced as a variable star by Friedrich and Schöffel (1971) and entered in the *New Catalogue of Suspected Variable Stars* (Kholopov *et al.* 1982) as NSV 12817. The star shows semi-regular pulsations with a predominant period of 50.93 days and several fading events during the coverage of ASAS-3 (Figure 18). Apart from several minor drops in brightness, which have been covered to various degrees, there is a major one with an amplitude of ~1.5 magnitude (V) around HJD 2454300. The symmetric declines and recoveries of the fadings are reminiscent of the behavior of DYPer stars (see, for example, Tisserand (2012)).

An object showing similar photometric variability is ASAS J095221-4329.8 (CD-42 5700), which entered the ACVS as a MISC-type variable. It is also characterized by several minor fadings and a major drop with an amplitude of about 2 magnitudes (V) at around HJD 2452200. Furthermore, the star exhibits semiregular pulsations, with a predominant period of ~65 days, which seem to continue during the obscuration phases (Figure 19).

Both objects show similar colors. The color indices of NSV 12817 (J–K = 1.20 (2MASS), B–V = 1.58 (APASS)) are indicative of a red star. ASAS J095221-4329.8 is classified as M6 in Tisserand *et al.* (2013b), which is in good agreement with its color indices of J–K = 1.29 (2MASS) and B–V = 1.78 (APASS).

The position of both stars in the near- and mid-infrared two-color diagrams is in accordance with that of the DYPer candidates of Alcock *et al.* (2001).

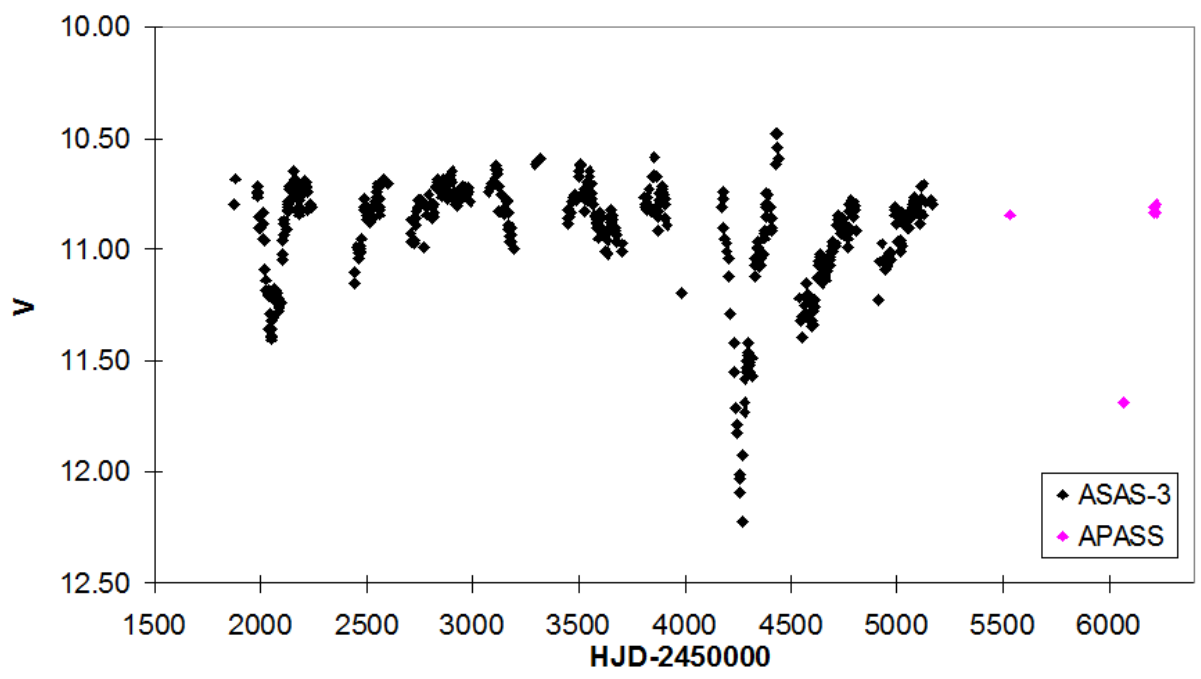

Figure 18. Light curve of NSV 12817, based on ASAS-3 and APASS data. Obvious outliers have been removed by visual inspection.

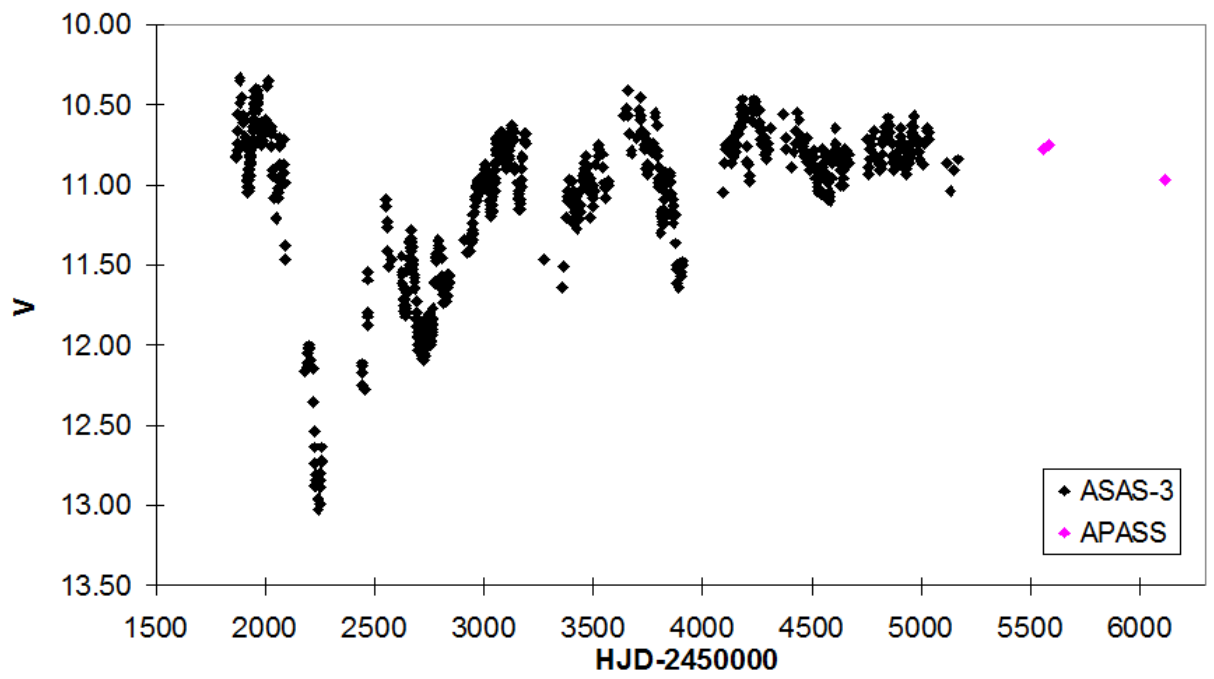

Figure 19. Light curve of ASAS J095221–4329.8, based on ASAS-3 and APASS data. Obvious outliers have been removed by visual inspection.

In fact, ASAS J095221-4329.8 was proposed as a possible DYPer variable by Hümmerich (2011) but rejected as such on spectroscopic grounds by Miller *et al.* (2012), who did not detect carbon compounds but strong titanium oxide bands (TiO), vanadium oxide (VO) and possibly hydrogen ("strong TiO, VO; H?"; see Miller *et al.* (2012), especially their Table 5). Both objects seem to be further examples of semiregular red giant stars that merit attention because of their significant fading events.

## 4. Acknowledgements

We thank the anonymous referee for valuable comments and suggestions that greatly improved the paper. We acknowledge with thanks the variable star observations from the AAVSO International Database contributed by observers worldwide and used in this research. This research has made use of the SIMBAD and VizieR databases operated at the Centre de Données Astronomiques (Strasbourg) in France. This work has also made use of EROS-2 data, which were kindly provided by the EROS collaboration. The EROS (Expérience pour la Recherche d'Objets Sombres) project was funded by the CEA and the IN2P3 and INSU CNRS institutes. Furthermore, this research has employed data products from the Two Micron All Sky Survey, which is a joint project of the University of Massachusetts and the Infrared Processing and Analysis Center/California Institute of Technology, funded by the National Aeronautics and Space Administration and the National Science Foundation, and the Wide-field Infrared Survey Explorer, which is a joint project of the University of California, Los Angeles, and the Jet Propulsion Laboratory/California Institute of Technology, funded by the National Aeronautics and Space Administration.